\def\cH{{\cal H}}
\def\ket#1{\left|\, #1\right\rangle}

\documentclass{article}
\usepackage[hypertex]{hyperref}

\title{Teleporting a qubit {\em \`a la\/} Everett}
\author{George Svetlichny\footnote{Departamento de Matem\'atica, Pontif\'{\i}cia Universidade Cat\'olica, Rio de Janeiro, Brazil \newline
svetlich@mat.puc-rio.br \hfill \url{http://www.mat.puc-rio.br/\~svetlich}}}
\begin{document}
\maketitle

\begin{abstract}If both Alice and Bob have access to a two-qubit ``background state" then,  by simulating Everett's many worlds interpretation of measurement, Alice can teleport a qubit to Bob, each  using  fixed unitaries. The Everett picture unifies unitaries, measurements, and classical communication into just unitaries, provided there are background states shared by all parties. We review some of the literature on this theme and point out an additional fact: If Bob has access to only one of the background qubits, teleportation is still possible by Alice sending Bob one classical bit of information gained from a measurement of the other qubit.
\end{abstract}

\section{Introduction}
The usual description of quantum-state teleportation  \cite{benn-etal:PRL70.1895} is given within the Copenhagen interpretation of quantum mechanics, and so involves projection measurements and classical communication.
 However, by now there are many versions of quantum mechanics without a separate measurement process. These include Bohmian mechanics  \cite{bddgz}, the consistent histories approach \cite{omnes}, Everett's many worlds \cite{dewgra}, its variant ``many minds" \cite{donald}, and any number of decoherence theories \cite{jzkgks}. One can ask how  teleportation proceeds  in these alternate formalisms. This should not only give new conceptual perspectives on what has been dubbed ``quantum information" but may also provide new mathematical tools to deal with quantum-state processing.  Not surprising, this theme has been taken up by various authors (Refs.  \cite{bras-brau-clev:PD120.43} --  \cite{vaid:PSA1994.1.211} and Ref.  \cite{Zeh}),  but the existence of this literature seems to be largely unknown\footnote{I was unaware of it when I had worked out the Everett teleportation scheme here presented, and a quick search on the internet did not bring up any other works.} and it's lessons unappreciated.

 We feel that both of the above mentioned benefits happen in the Everett version. There is a trade-off between quantum and classical communication, and the various types of actions of the parties get a unified description.

We shall first present an Everett description of teleportation based on the usual incorporation into unitary dynamics of the von Neumann measurement paradigm, and follow this by some remarks on the cited literature.

\section{The Everett teleport}
Recall first the usual teleportation protocol. Alice has an unknown qubit \(\phi_u=\alpha\ket0_u+\beta\ket1_u\) which she wants to send to Bob. We use subscripts on kets to also label the Hilbert spaces to which they belong. Thus \(\phi_u\in \cH_u\).
Each one of the parties also has one of a pair of entangled qubits in the  state  \[\Lambda=\ket0_a\ket0_b+\ket1_a\ket1_b.\]
Here and below, states are not assumed to be normalized.

In the usual teleportation scheme, Alice measures her pair of qubits in the Bell basis:
\begin{displaymath}
\psi_{xy}=\ket x\ket y+(-1)^y\ket{x+1}\ket{y+1}
\end{displaymath}
where \(x,\,y\) are qubit indices, that is \(0\) or \(1\), and all sums are modulo \(2\).
Using two classical bits (c-bits) Alice now sends Bob the result of her measurement, say the two numbers \(x\) and \(y\) if the measurement projected her pair of qubits onto \(\psi_{xy}\). Bob then applies  the  unitary transformation
\begin{equation}\label{bobu}
\ket z_b\mapsto (-1)^{y(z+1)}\ket{z+x+y}_b
\end{equation} to his partner of \(\Lambda\), which is now disentangled from the rest,  producing the state \(\phi_b=\alpha\ket0_b+\beta\ket1_b\), and exact replica of \(\phi_u\).

Recall now the von Neumann description of a measurement process. For simplicity assume all Hilbert spaces finite dimensional. We have a system \(S\) described by vectors in a Hilbert space \(\cH_S\) and we wish to measure an observable \(A\) with non-degenerate eigenbasis \(\ket i_S,\,i=1,\dots,N\). We also have a measuring apparatus that corresponds to the above observable and which is described by vectors in a Hilbert space \(\cH_A\) with a set of ``pointer states" \(\ket i_A\,i=1,\dots,N\). We assume that \(\ket 1_A\) is also the initial state of the apparatus before any measurement takes place. The measurement process then is described by a unitary transformation \(U\) for which
\begin{displaymath}
U:\ket i_S\ket 1_A\mapsto \ket i_S\ket i_A
\end{displaymath}
holds. In other words, the pointer points to \(i\) if the state of the system is in \(\ket i_S\). Of course, since \(U\) is linear one has for a superposition of eigenstates of the system:
\begin{displaymath}
U:\left(\sum_ic_i\ket i_S\right)\ket 1_A\mapsto \sum_ic_i\ket i_S\ket i_A.
\end{displaymath}

The right hand side is a an entangles state of system and apparatus without well defined pointer positions. Since we don't experience this in practice, we have the by now much vented ``measurement problem" for which there are countless solution proposals. In the Everett scheme the right hand side corresponds to multiple worlds, one for each term with \(c_i\neq0\), into which the world has split and in each one of which  a definite pointer position is perceived by observers. We do not argue for this ontology, but proceed with the mathematical analysis.  Thus we now model Alice's measuring apparatus by a two-qubit  Hilbert space \(\cH_E=\cH_c\otimes\cH_d\) with basis \(\ket {xy}_E=\ket x_c\ket y_d\) which are the ``pointer positions" for the Bell basis. We use \(\ket {00}_E\) as the initial state. Alice's measurement process is then described by a unitary which satisfies
\begin{displaymath}
U:\ket {00}_E\otimes\psi_{xy} \mapsto \ket {xy}_E\otimes\psi_{xy}
\end{displaymath}
and which can be extended to the twelve-dimensional orthogonal complement in an arbitrary fashion.

We further assume that  Alice and Bob {\em  both\/} have access to \(\cH_E\), which is to be considered  a common ``background system".

The result of applying \(U\), acting on the subproduct \(\cH_E\otimes \cH_u\otimes \cH_a\), to \(\ket0_c\ket0_d\otimes\phi_u\otimes \Lambda\)  is:
\begin{eqnarray}\nonumber
&\ket{00}_E\otimes\psi_{00}\otimes\{\alpha\ket0_b+\beta\ket1_b\} + \ket{01}_E\otimes\psi_{01}\otimes\{\alpha\ket1_b-\beta\ket0_b\}\,+&\\ \label{egoo} &\ket{10}_E\otimes\psi_{10}\otimes\{\alpha\ket1_b+\beta\ket0_b\} + \ket{11}_E\otimes\psi_{11}\otimes\{-\alpha\ket0_b+\beta\ket1_b\}.&
\end{eqnarray}

Bob now performs a tripartite unitary operation \(V\) acting on \(\cH_E\otimes\cH_b\)
defined by
\[V:\ket{xy}_E\ket z_b\mapsto(-1)^{y(z+1)}\ket{xy}_E\ket{z+x+y}_b.\]
Compare with (\ref{bobu}).
The result of VU applied to the initial state is
\[\left(\sum_{xy}\ket{xy}_E\otimes\psi_{xy}\right)\otimes\phi_b\]
where \(\phi_b=\alpha\ket0_b+\beta\ket1_b\). In other words Bob gets an exact copy of Alice's state \(\phi_u\) disentangled from the rest.

We have thus succeeded in teleporting a qubit using two fixed unitaries, provided both parties have access to a background state which performs the role of measurement and classical communication. From the practical side, this is arguably no progress at all, for if Alice and Bob are far apart, no such background states are available. Conceptually and mathematically we have succeeded in unifying all the allowed actions of the parties, unitary transformations, measurement, and classical communication. They are all implemented by unitary transformations in an Everett universe. The teleportation act described above can be thought of as happening within an ``Everett bubble" in an otherwise Copenhagen universe. The entangled state \(\Lambda\) and the initial neutral state of the ``instrument", \(\ket{00}_E\), should be considered to be  prepared in the usual Copenhagen way through measurement. The state \(\phi_b\) that Bob finally has can again be thought of as entering a Copenhagen universe, as it is totally disentangled from everything in the ``bubble".

The boundary between the ``bubble" and the surrounding Copenhagen universe is somewhat flexible. We have succeeded above in substituting classical communication of two c-bits by a sharing of two qubits of the environment. If Bob has access to only one of these qubits, say the one in \(\cH_c\)  he can still make use of this since Alice can now sends him {\em  one\/} c-bit of information gathered from the qubit to which Bob has no access and again teleport \(\phi_u\) to Bob. This half-classical-half-quantum communication can take place in two possible orders, first Alice acts then Bob, or vice versa. Let thus Alice first perform a measurement on state (\ref{egoo}) of the qubit in \(\cH_c\) using the basis \(\ket0,\,\ket1\) with outcomes \(0\) and \(1\) respectively. Denote Alice's outcome by \(z\). The resulting projected state is the first line of expression (\ref{egoo}) if the outcome is \(0\) and is the second line if it is \(1\). Alice now transmits the result of her measurement (one c-bit's worth) to Bob.
Depending on \(z\),  Bob applies the following unitary on \(\cH_d\otimes\cH_b\)
\begin{displaymath}
W_z:\ket x_d\ket y_b\mapsto (-1)^{x(y+1)}\ket x_d\ket{x+y+z}_b
\end{displaymath}
which results in a final state
\begin{equation}\label{bobhh}
\left\{\ket{z0}_E\otimes\psi_{z0}+\ket{z1}_E\otimes\psi_{z1}\right\}\otimes \phi_b.
\end{equation}
Thus in the end, for both values of \(z\), Bob holds a qubit disentangled from the rest which is a duplicate of Alice's qubit \(\phi_u\).

Consider now the opposite order. Bob now applies \(W_0\) to  state (\ref{egoo})
 which results in
\begin{eqnarray}\nonumber
& \left\{\ket{00}_E\otimes \psi_{00}+\ket{01}_E\otimes\psi_{01}\right\}\otimes \{\alpha\ket0_b+\beta\ket1_b\}\,+& \\ \label{eegoo}
&\left\{\ket{10}_E\otimes \psi_{10}+\ket{11}_E\otimes \psi_{11}\right\}\otimes \{\alpha\ket1_b+\beta\ket0_b\}.\,\,\,&
\end{eqnarray}

Alice now measures the qubit in \(\cH_c\) as before. If the result is \(0\), the projected state is the first line in expression (\ref{eegoo}) and if it is \(1\), it's the second line. As before, Alice send her result \(z\) to Bob who acts with the unitary \(\ket x_b\mapsto \ket{x+z}_b\) in \(\cH_b\). Again, for both cases of \(z\),  the final state is (\ref{bobhh}), and so there's a state \(\phi_b\) in \(\cH_b\) disentangled from the rest which is a perfect copy of Alice's \(\phi_u\).

One sees here a satisfying equivalence between the use of one qubit in the ``Everett bubble" and one c-bit in the Copenhagen exterior.

The Everett interpretation still raises many conceptual issues but one can be assured of its mathematical consistency. It is then plausible that many, if not all, instances of quantum state processing utilizing unitaries, measurements and classical communication can be equivalently described solely by unitaries in appropriate ``Everett bubbles" by introducing universally available ``background states". This also accords with the view  that there really is no classical regime and everything is truly quantum. Classicity would be an emergent appearance and all processes would in the end be implemented by unitaries provided enough quantum degrees of freedom are included in the description.

\section{Remarks}

We now briefly describe the cited literature. Our aim is not to give a critical analysis of the content, but only to provide some introductory comments for the interested reader.

Teleportation with just unitaries is described in many  of the references. Brassard et al.\ \cite{bras-brau-clev:PD120.43} presents it in terms of quantum computation primitives, CNOT and Hadamard gates to be precise. Mermin  \cite{merm:PRA65.012320} describes  essentially the same quantum circuit. His aim however is to show how such a circuit could be extracted in a transparent way from a classical teleportation circuit and also generalized to q-dit systems. Busch \emph{et al.} \cite{busc-etal:PLA284.141} teleport a qubit with  essentially the same two unitaries presented here. Their intent is to "avoid any reference to the postulate of the state collapse", an evident Everettian attitude, though no mention of Everett is made. Timpson in Appendix B of his paper \cite{timp:quant-ph/0509048} presents a scheme which, except for differences of notation, is identical to the one presented here. He also discusses in greater detail and in philosophical terms how teleportation works from the Everett and the Bohmian points of view. In this discussion classical communication is not explicitly treated in terms of the underlying processes (unitary evolution and Bohmian trajectories respectively) of the two interpretations, however the author's intent is that they be so understood\footnote{Private communication.}, and in fact the scheme in his Appendix B does subsume communication into unitaries. Joos  \cite{joos} also presents a scheme in essence identical to the one here.

The paper of Motoyoshi \emph{et al.}  \cite{moto-etal:PTP97.819} although entitled ``A set of Conditions for Teleportation without Resort to von Neumann's Projection Postulate" is not a true Everettian interpretation of teleportation. The paper is founded  on the claim that the original teleportation scheme of Bennett \emph{et al.} \cite{benn-etal:PRL70.1895} is ``insufficient due to the unclear treatment of the EPR", and this in turn is based on  alternative views of the process of measurement and conservation laws therein, and this view is not Everettian.

Ralph  \cite{ralp:OL24.348} proposes an all-optical teleprotation scheme in principle not involving any irreversible processes, mentioning Braustein  \cite{brau:PRA53.1900} in this regard.

Deutsch and Hayden  \cite{deut-hayd:PRSLA456.1759} present a ``Heisenberg picture" version of teleportation. Such a picture, introduced by Gottesman  \cite{gott:quant-ph/9807006}, may be viewed as a new measurement-less interpretation of the quantum formalism. The authors' aim is to argue for ``locality of quantum information". A critical appraisal of this argument  is provided by  Timpson  \cite{timp:FOP35.313}.

Lev Vaidman  \cite{vaid:PSA1994.1.211} argues that the Everett interpretation removes some of the paradoxical aspects of both interaction-free measurements  \cite{elit-vaid:FOP23.987} and teleportation. Concerning teleportation this can be summed up in the quote ``the teleportation procedure does not move the quantum state: the state was, in some sense, in the remote location from the beginning". Joos  \cite{joos} quotes the same passage, stating that this may well be a proper characterization of the teleportation situation. This is also in accord with our own view that entanglement is unlikely a ``channel for information flow" as is often stated, but, along with classical information,  defines conditions for quantum state availability  \cite{svet:quant-ph/0601093}.

Zeh  \cite{Zeh} also argues, from a general decoherence perspective, for a similar view:   the nonlocal entangled state is such that it ``contains, in one of its components, precisely what is later claimed to be ported already at its final position".

It is likely that other literature exists that is relevant to the present theme, but unfortunately any such would be scattered and not easily identified.

We see from all these works that looking at teleportation from various interpretational perspectives has certainly enriched our understanding of quantum phenomena and structures. There could also be technical benefits: the unifying power of the Everett interpretation, using just unitaries, could in principle provide considerable simplification in mathematical analyses of quantum-state processing. This is certainly worth investigating.

\section{Acknowledgements}
I thank professors Cristopher~G.~Timpson, Lev~Vaidman and H.~Dieter~Zeh for acquainting me with previous literature on the subject and for very instructive comments. This research  was partially financed by the Conselho Nacional de Desenvolvimento Cient\'{\i}fico e Tecnol\'ogico (CNPq).

\end{document}